# Superconductivity at 7.4 K in Few Layer Graphene by Li-intercalation

**Anand P. Tiwari**[1+], **Soohyeon Shin**[2+], **Eunhee Hwang**[1], **Soon-Gil Jung**[2], **Tuson Park**[2]* and **Hyoyoung Lee**[1]*

1. Centre for Integrated Nanostructure Physics (CINAP), Institute of Basic Science (IBS), Department of Chemistry, Sungkyunkwan University, 300 Cheoncheon-Dong, Jangan-Gu, Suwon, Gyeonggi-Do 440-746, South Korea

2. Department of Physics, Sungkyunkwan University, Suwon 440-746, South Korea

   **\* Corresponding Author**:

   **Prof. Hyoyoung Lee**
   SKKU-Fellow, Professor of Chemistry, Sungkyunkwan University, Suwon 440-746, Korea.
   Tel: +82-31-299-4566; Fax: +81-31-290-5934
   E-mail: hyoyoung@skku.edu,

   **Prof. Tuson Park**
   Professor of Physics, Sungkyunkwan University, Suwon 440-746, Korea.
   Tel: +82-31-299-4543
   E-mail: tp8701@skku.edu,

   + Equally Contributed

**Abstract**

**Superconductivity in graphene has been highly sought after for its promise in various device applications and for general scientific interest. Ironically, the simple electronic structure of graphene, which is responsible for novel quantum phenomena, hinders the emergence of superconductivity. Theory predicts that doping the surface of the graphene effectively alters the electronic structure, thus promoting propensity towards Cooper pair instability [1, 2]. Here we report the emergence of superconductivity at 7.4 K in Li-intercalated few-layer-graphene (FLG). The absence of superconductivity in three-dimensional Li-doped graphite underlines that superconductivity in Li-FLG arises from the novel electronic properties of the two-dimensional graphene layer. These results are expected to guide future research on graphene-based superconductivity, both in theory and experiments. In addition, easy control of the Li-doping process holds promise for various device applications.**

The discovery of proximity-induced superconductivity in graphene has stimulated considerable pursuit for intrinsic superconductivity [3]. Similar to bulk graphite, graphene itself is not superconducting because of the vanishingly small electronic density of states at the Fermi level, requiring modification of its electronic structure in order to facilitate superconductivity [1, 2]. Several efforts have been initiated to introduce a change in the band structure by depositing atoms or molecules on the graphene and through the application of an external electric field [4-9]. Recently, the intercalation of guest atoms into graphene layers in carbon allotropes has been reported to induce superconductivity [10-12], though which electronic states, intercalant- or graphene-derived, and which phonons are responsible for the superconducting Cooper pairing remains unclear [13-17].

In three-dimensional (3D) graphite intercalated compounds which utilize alkalis and alkaline earths as the intercalant, the distance *h* between the intercalant and the graphene layers plays a critical role in superconductivity: critical temperatures ($T_c$s) for $BaC_6$ (*h*=2.62 Å), $SrC_6$ (=2.47 Å), and $CaC_6$ (=2.26 Å) are 0.064, 1.65, and 11.5 K, respectively [18] (see Fig. S1, Supplementary Information). When the distance between the intercalated atom and graphene plane is small, the deformation potential as well as carbon out-of-plane modes is to be large, resulting in enhanced electron-phonon coupling and higher $T_c$s [19, 20]. However, when the intercalant distance *h* is too small, as in $LiC_6$ graphite (h=1.85 Å), the superconductivity can be completely destroyed because a strong confinement of the interlayer state in a narrow region shifts the intercalant band well above the Fermi energy [14]. A band structure calculation predicts that the empty interlayer state can be returned to the Fermi level by removing the quantum confinement [1]. In the 2D graphene $LiC_6$, where the quantum confinement is removed, the electron-phonon coupling from the low-energy lithium modes and carbon out-of-plane vibrations is predicted to be strong enough to induce superconductivity at as high as 8.1 K.

Here we report the discovery of the superconducting (SC) phase below 7.4 K in the Li-intercalated few-layer-graphene (FLG). Zero-field-cooled (ZFC) magnetization (*M*) measurements of the Li-FLG reveal a sharp suppression below 7.4 K due to the Meissner effect, where $T_c$ is progressively suppressed with increasing magnetic field. Magnetization hysteresis loop as a function of magnetic field, which arises due to the trapped magnetic flux lines, becomes more pronounced with decreasing temperature. The upper critical field $H_{c2}$ from the *M-T* measurements is 1538 Oe and the lower critical field $H_{c1}$ is 124 Oe, indicating that the Li-FLG is a prototypical type II superconductor with the Ginzburg-Landau parameter $\kappa$ ($=\lambda/\xi$) of 3.52. Here, $\lambda$ and $\xi$ is the penetration depth and SC coherence length, respectively. Observation of $T_c$ at 7.4 K in the graphene-based LiC$_6$ and its good agreement with the predicted $T_c$ indicate that the intercalant-derived band and the carbon out-of-plane vibrations are important to forming the SC Cooper pairs [1].

Li-intercalated few-layer graphene (FLG) samples are synthesized via a solution reaction of graphite flakes and Li, through several successive steps. All reactions for sample preparation are performed in a glove box with purified argon atmosphere. Natural graphite flakes (Alfa Aesar 99.9%), with a mean diameter of 1 mm, and Li (Sigma Aldrich) are taken for the reaction. First, the graphite flakes are placed in a round-bottom flask; Li is added and then degassed in a vacuum while heating. The resulting mixtures are raised to 200 $^0$C for 24 hrs. Then, 150 ml 1,2-Dimethoxyethane (99.9%, Sigma Aldrich) is added in a freshly-prepared alloy of graphite and Li, and the Li-intercalated graphene is dispersed via ultrasonication for 24 hrs. The reaction mixture solution is then stirred for 7 days at room temperature. After standing for 24 hrs, the solution displays stratification, the upper clear solution can be poured off and the medial part of the solution evaporated. The sample must be kept in an argon atmosphere for 24 hrs to enhance the stability of Li between the layers of few-layer graphene.

Figure 1a compares XRD patterns of the as-synthesized, Li-deposited few-layer graphene (Li-FLG) and pristine graphite in the top and bottom parts, respectively. The (002) reflection of the graphite precursor at 26.4 ° indicates that the material is AB stacked [21]. The width of the (002) reflection of the Li-FLG is significantly broadened due to the Li intercalation and the peak position slightly moved to a smaller angle of 26.11 °, indicating that the interlayer spacing of the graphene layers is almost same as that of the graphite. Atomic force microscopy (AFM) image displays the Li-intercalated graphene flakes (bright spots) on the silicon surface where the dimension is approximately 30 x 20 x 2 nm$^3$ (see Fig. 1b). The line profile of the Li-FLG, as shown in Fig. 1c, shows that the average height of the flakes is ~2 nm, implying that 5 to 6 layers of graphene are formed on the substrate [22]. X-ray photoelectron spectroscopy (XPS) analysis (see Fig. S2, Supplementary Information) shows that 3.09 atomic % of Li atoms were intercalated in the graphene flakes.

Magnetization of Li-intercalated graphene at 20 Oe is shown as a function of temperature in Fig. 2a. Both zero-field-cooled (ZFC) and field-cooled (FC) data show a drop below 7.4 K, which manifests the Meissner effect of magnetic flux expulsion below the superconducting (SC) phase transition temperature ($=T_c$). The FC magnetization at 20 Oe drops slightly at the $T_c$ onset and gradually increases with decreasing temperature, indicating the presence of local spins that may be introduced from such various defects on the graphene layers as vacancies, frustration, or hydrogen chemisorption [23-25]. The anomalous low-temperature upturn in magnetization can be described by the Curie-Weiss law, where the concentration of magnetic defects is estimated to be 0.012 % with an assumption that the size of moment for each defect is 2 $\mu_B$ [26, 27] (see Fig. S3, Supplementary Information).

Figure 2b displays the evolution of the superconducting transition temperature and the SC shielding fraction $4\pi$ ($\chi_{ZFC} - \chi_{bkg}$) for several magnetic fields. Since the FC magnetic susceptibility $\chi_{FC}$ has a negligible change near $T_c$ from the Curie-Weiss background (see Fig. S4, Supplementary Information), it is approximated as the background $\chi_{bkg}$, the magnetic susceptibility of the Li-FLG in the normal state. With increasing magnetic field, $T_c$ is gradually suppressed. The SC shielding fraction is approximately 0.06% for the pellet of Li-FLG flakes, which reflects a small fraction of Li atoms that are successfully intercalated in the graphene flakes (see Fig. S2, Supplementary Information). When the size of the SC grains is comparable to the London penetration depth, small SC volume fraction is often reported in the powder form of the specimen because the magnetic flux expulsion arises from the aggregated SC grains: the SC shielding fraction was reported to be 0.05% for graphite-sulfur (C-S) composites [28, 29]

The dependence on magnetic field of magnetization $M(H)$ of Li-doped FLG is displayed in Fig. 3a, where $\Delta M$ is the magnetization after subtracting $M$ at 10 K in the normal state of Li-FLG, i.e., $\Delta M = M(H, T) - M(H, 10K))$ (see Fig. S5, Supplementary Information). At low fields, $\Delta M$ at 2 K decreases linearly with increasing $H$ due to the Meissner effect, deviates from the linear dependence above $H_{c1}$ (which is marked as an arrow in the inset to Fig. 3a), and starts to increase with further increasing magnetic field due to vortices in the mixed phase of the type II superconductor. At higher fields, $\Delta M$ crosses zero to a positive value and increases with field. These peculiar $M$-$H$ hysteretic loops below $T_c$ can be explained by the superposition of the SC diamagnetic component and the paramagnetic contribution from the local spins that caused the anomalous upturn in the temperature dependence of $M$ below $T_c$ (see Fig. 2a). When temperature is raised close to $T_c$ of 7.4 K, the hysteretic behaviour in the $M$-$H$ curve is almost negligible.

Figure 3b describes the temperature dependence of the upper critical field, $H_{c2}$, which was obtained from the $M$-$T$ measurements. The orbital depairing field of Li-FLG is estimated to be 1538 Oe by using the Werthamer–Helfand–Hohenberg (WHH) model for the dirty type II superconductor [30], where the slope at $T_c$ is -304 Oe/K: $H_{c2}(T=0) = -0.69T_c(dH_{c2}/dT)_{T=Tc}$. The Ginzburg-Landau coherence length estimated from the relation $H_{c2} \sim \Phi_0/2\pi\xi^2$ is $\xi$=462 Å, where $\Phi_0(=2.07\times10^{-15}\,T\cdot m^2)$ is a flux quantum. The lower critical field $H_{c1}$, where magnetic flux begins to penetrate the Li-FLG, is multiplied by a factor of 2 for comparison, and is plotted as a function of temperature in Fig. 3b. $H_{c1}$ initially increases with decreasing temperature and saturates to 124 Oe at lower temperatures. When estimated from $H_{c1}$ ($\sim\Phi_0/2\pi\lambda^2$), the penetration depth $\lambda$ by which the applied field extends into the SC state is estimated to be 1628 Å. The Ginzburg-Landau parameter, κ (=$\lambda/\xi$), is 3.52, indicating that the Li-FLG is a prototypical type II superconductor.

Electrical resistance measurements, $R(T)$, were performed on a pellet of Li-FLG flakes, where the average dimension of the flakes is 30 x 20 x 2 $nm^3$. Figure 4a representatively shows the zero-field resistance of the Li-FLG as a function of temperature. $R(T)$ increases gradually with lowering temperature, but deviates from the linear increase, showing a strong enhancement below 7.4 K, the SC transition temperature determined from the $M$-$T$ measurements. When the zero-field resistance was subtracted by the resistance for 2 kOe (>$H_{c2}$), $\Delta R=R(0\ kOe)-R(2\ kOe)$, the anomalous enhancement is clearly distinguishable and reproducible in the measured pellets of the Li-FLG flakes, as shown in Fig. 4b and 4c. The anomalous increase in $R$ below $T_c$ is similarly observed in two-dimensional superconductors that are in the regime of weakly localized Cooper pairs [31]. When combined with the small SC shielding fraction, the anomalous upturn in $R$ indicates that superconductivity is localized within the Li-FLG grains or islands. Future study on the nano-scale transport measurements such as scanning tunneling microscopy (STM) and magnetic force microscopy (MFM) is expected to provide direct information on the SC properties of the superconducting Li-FLG flake.

Observation of superconductivity in the Li-intercalated FLG flakes implies that the removal of the quantum confinement of the intercalated Li state is a key to superconductivity where the deconfinement returns the empty interlayer state across the Fermi level and enhances the electron-phonon coupling that was negligible in the bulk Li-graphite. The SC transition temperature 7.4 K is slightly lower than the predicted 8.1 K for the Li-intercalated single graphene layer [1], which could be ascribed to the magnetic defects that caused the unusual upturn in the *M-T* measurements and the paramagnetic background in the *M-H* hysteresis loop.

To summarize, we reported the superconductivity at 7.4 K for the Li-FLG, which is the highest $T_c$ among the intercalant FLG compounds. The discovery not only confirms the theoretical prediction that the two-dimensional graphene-based superconductivity is very different from the bulk graphite based counterpart, but also is expected to expedite further research on superconductivity in low-dimensional materials, particularly aiding in the ability to achieve higher $T_c$s through the manipulating layer thickness and the adsorption process: 17~18 K of $T_c$ is theoretically predicted in the graphene-based $Li_2C_6$ compound [1]. Successful synthesis of SC Li-FLG through wet chemistry holds promise for nanoscience applications because of its simple control afforded over the Li-doping process.

**Materials and Methods:**

*Characterizations* The sample is washed with ethanol for several minutes to remove the absorbed Lithium on the surface of the FLG. The resulting sample is then placed in a vacuum chamber overnight to remove the residual ethanol and homogenize the distribution of Lithium. The structural and microstructural characterizations of the sample are studied via X-ray Diffraction (XRD) (Rigaku Ultima IV), X-ray photoelectron spectroscopy (ESCA 2000, VG Microtech), Atomic Force Microscopy (AFM) (Agilent 5100 AFM/SPM system), and Transmission Electron Microscopy (TEM) (JEOL JEM-2100F). Magnetic properties, such as the temperature dependences of magnetization (*M*) and magnetic hysteresis (*M - H*) loops, for a pellet of Li-doped FLG flakes are investigated via a magnetic property measurement system (MPMS) Quantum Design).

**Acknowledgements:** We thank S. Oh for helping magnetization measurements. This work was supported by the Creative Research Initiatives funded by the Ministry of Science, ICT and Future Planning (No. IBS-R011-D1 & 2012R1A3A2048816).

**Author Contributions:** APT and SS contributed equally to this work. All authors discussed the results and commented on the manuscript. APT synthesized the samples. EH performed the structural characterizations. SS and SGJ performed and analysed MPMS experiments. APT, SS, TP and HL wrote the manuscript.

**Figure Legends.**

**Figure 1** Structural characterization of the Li-intercalated few-layer-graphene (FLG). **(a)** X-ray diffraction patterns of Li–intercalated FLG and graphite are plotted as a function of the diffraction angle 2θ in the top (blue line) and bottom (black line) parts, respectively. **(b)** Surface

image of Atomic Force Microscopy and **(c)** its corresponding line profile of as-synthesized Li-FLG.

**Figure 2** Magnetic characterization of Li-FLG. **(a)** Zero-field-cooled (ZFC) and field-cooled (FC) magnetization for a pellet of Li-FLG flakes is plotted as a function of temperature for 20 Oe. The onset of the SC phase transition is marked by an arrow. **(b)** The superconducting shielding fraction of the Li-FLG is estimated for 20 (squares), 100 (cicles), 200 (up triangles), 300 (down triangles), 500 (diamonds), and 1000 Oe (hairs). Here the background was approximated as the FC susceptibility because there was negligible change at $T_c$ in the FC magnetization. Arrows indicate the onset of the Meissner effect for each field.

**Figure 3 (a)** Subtracted magnetization $\Delta M$ is plotted as a function of magnetic field at 2 (squares), 4 (circles) and 6 K (triangles), where the magnetization $M(H)$ at 10 K is used as a reference in the normal state: $\Delta M = M - M(10\ K)$. Inset is a blow-up of the 2 K hysteresis loop near zero field, and the arrow marks the lower critical field, $H_{c1}$, where magnetic flux begins to penetrate the superconducting Li-FLG. (b) Temperature-magnetic field phase diagram of the superconducting Li-FLG. Upper critical field, $H_{c2}$ (squares), and lower critical field, $H_{c1}$ (circles), are plotted as a function of temperature. For comparison, $H_{c1}$ was magnified by a factor of 2.

**Figure 4** Electrical resistance of Li-FLG. **(a)** Electrical resistance of Li-FLG is plotted as a function of temperature for sample #1. The zero-field electrical resistance is subtracted by the normal-state resistance for 2 kOe (> $H_{c2}$), i.e. $\Delta R = R(0\ Oe) - R(2\ kOe)$, and plotted in **(b)** and **(c)** for sample #1 and #2, respectively. Solid lines are guides to eyes and arrows mark the $T_c$ determined from *M-T* measurements.

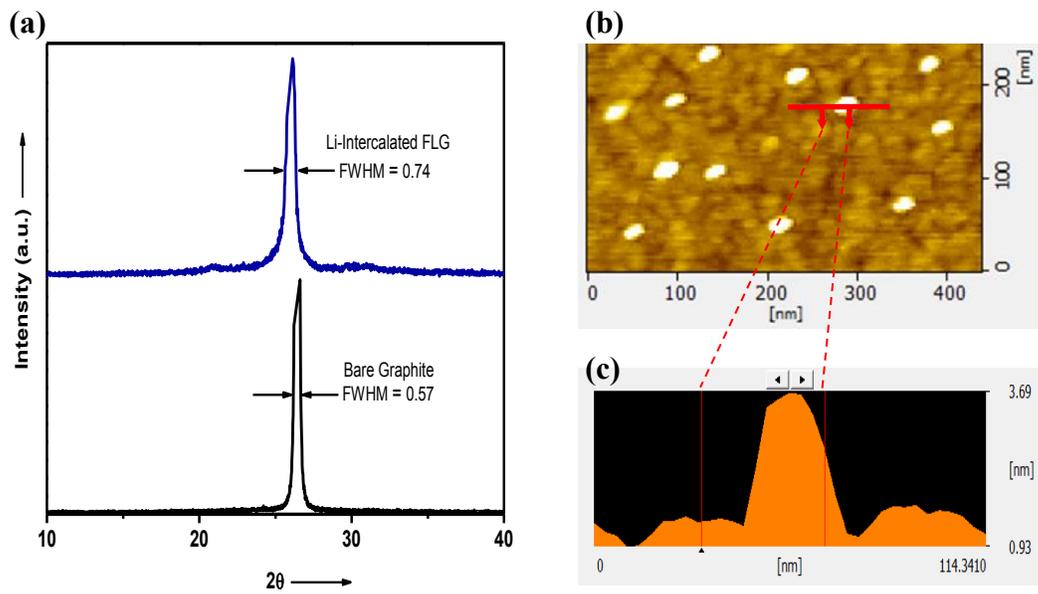

Figure 1

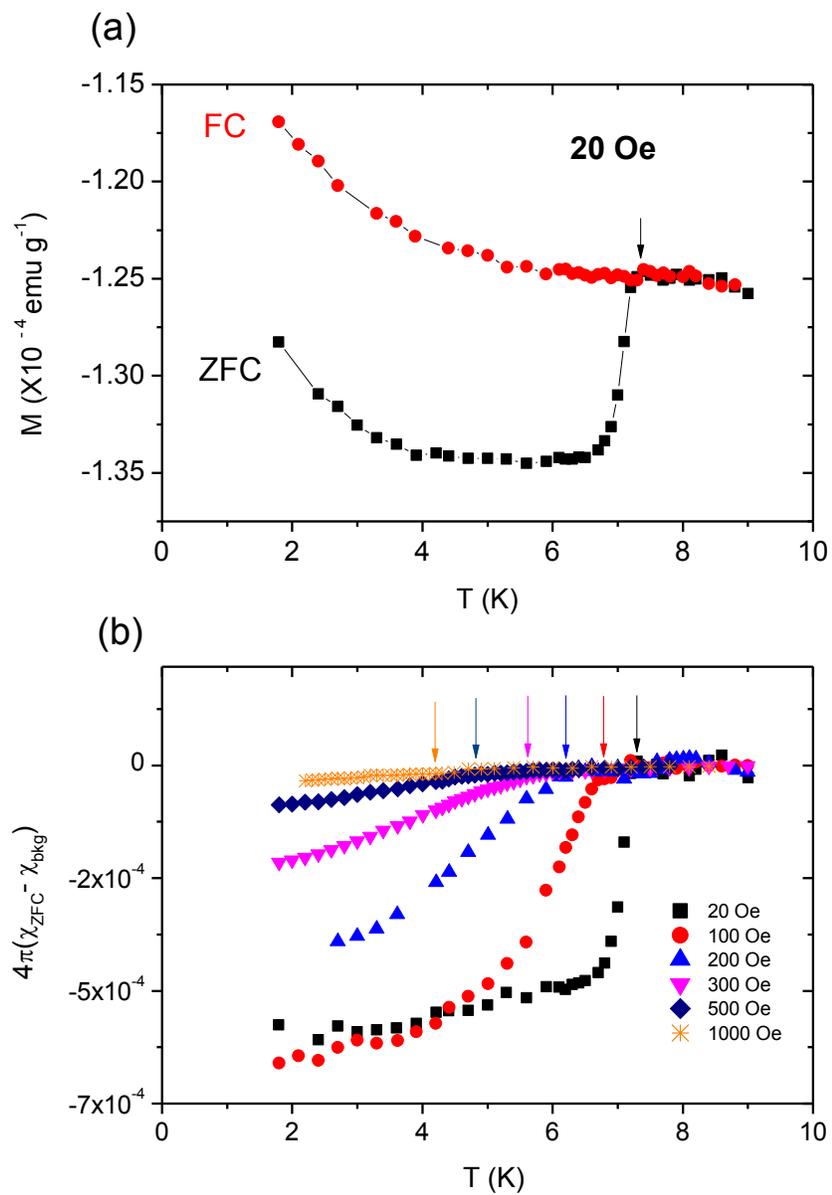

Figure 2

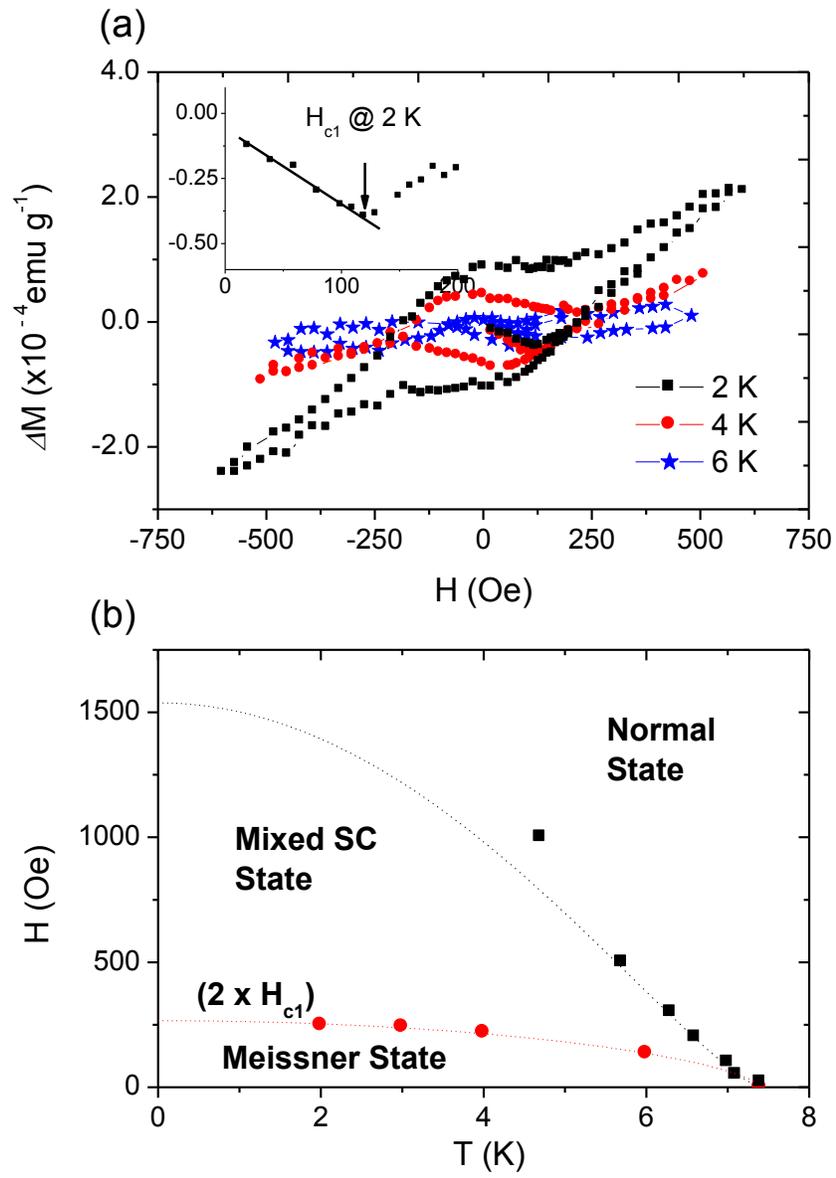

Figure 3

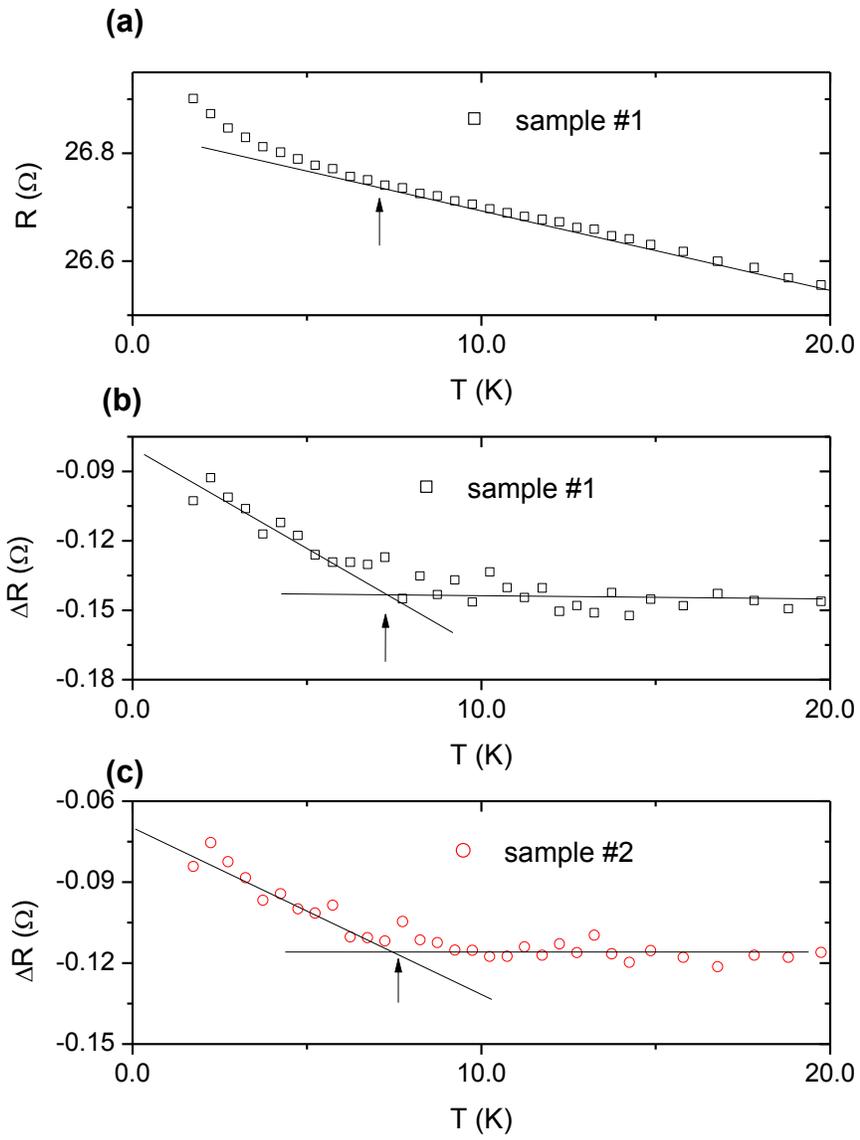

Figure 4

# Supplementary Information for

# Superconductivity at 7.4 K in Few Layer Graphene by Li-intercalation

Anand P. Tiwari[1+], Soohyeon Shin[2+], Eunhee Hwang[1], Soon-Gil Jung[2], Tuson Park[2]* and Hyoyoung Lee[1]*

1. Centre for Integrated Nanostructure Physics (CINAP), Institute of Basic Science (IBS), Department of Chemistry, Sungkyunkwan University, 300 Cheoncheon-Dong, Jangan-Gu, Suwon, Gyeonggi-Do 440-746, South Korea

2. Department of Physics, Sungkyunkwan University, Suwon 440-746, South Korea

In this Supplement, we present figures that support results in the main text. The list is

Figure S1. $T_c$ vs interlayer distance (Å) for graphite-based and graphene-based $MC_6$ compounds (M= Ca, Yb, Sr, Ba, Li).

Figure S2. X-ray photoelectron spectroscopy analysis of Li-FLG.

Figure S3. Molar susceptibility of Li-FLG due to paramagnetic impurities.

Figure S4. *M vs T* of Li-intercalated FLG for several magnetic fields

Figure S5. Field dependence of the magnetization of the Li intercalated FLG.

Supplementary References.

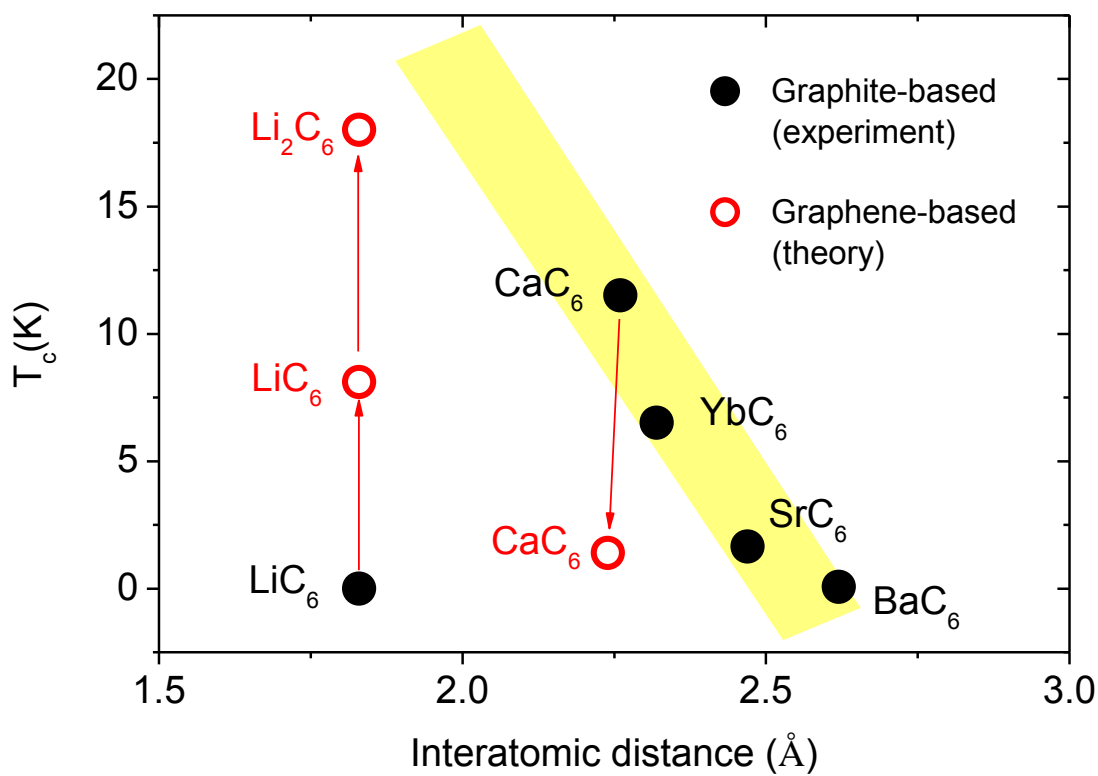

**Figure S1. $T_c$ vs interlayer distance** (Å) for graphite-based (solid circles) and graphene-based (open circles) $MC_6$ compounds (M= Ca, Yb, Sr, Ba, Li). $T_c$s for the graphite-based SCs are obtained from the experimental reports [S1-S4], while $T_c$s for the graphene-based SCs are predicted by a theory [S5]. We note that $T_c$s for the graphene-based $LiC_6$ and $Li_2C_6$ are predicted to be 8 and 18 K, respectively, while superconductivity has yet to be reported in the graphite-based $LiC_6$. This is opposite of $CaC_6$, where $T_c$ for the graphene-based compound is predicted to decrease to 1.4 K from the 11.5 K for the graphite based counterpart. The yellow shade is a guide for the eyes.

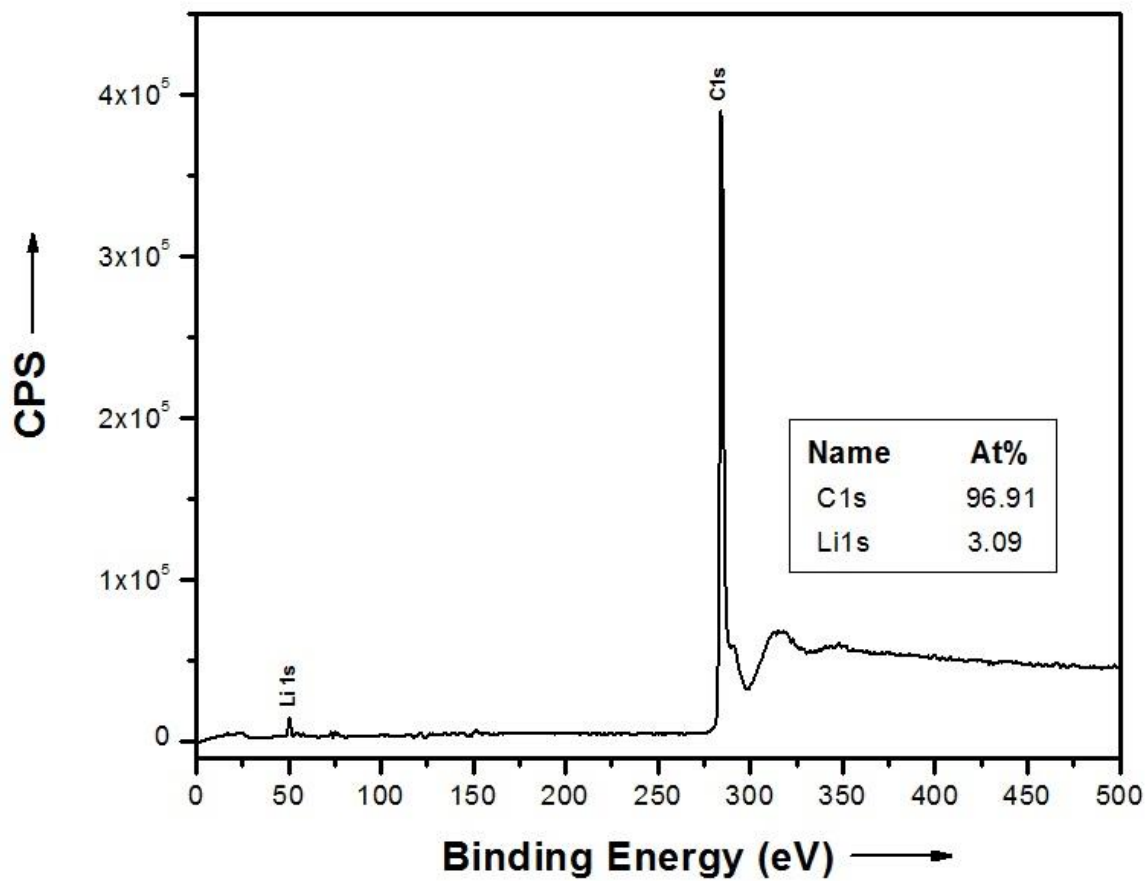

**Figure S2. XPS analysis of Li-FLG.** To determine the atomic composition of Li-intercalated few layer graphene X-ray photoelectron spectroscopy (XPS) measurements were carried out in the region of 0 ~500 eV. Figures S1 shows that the sample contains the C and Li elements and the atomic ratio of the elements are summarized in the insert of Figure S1.

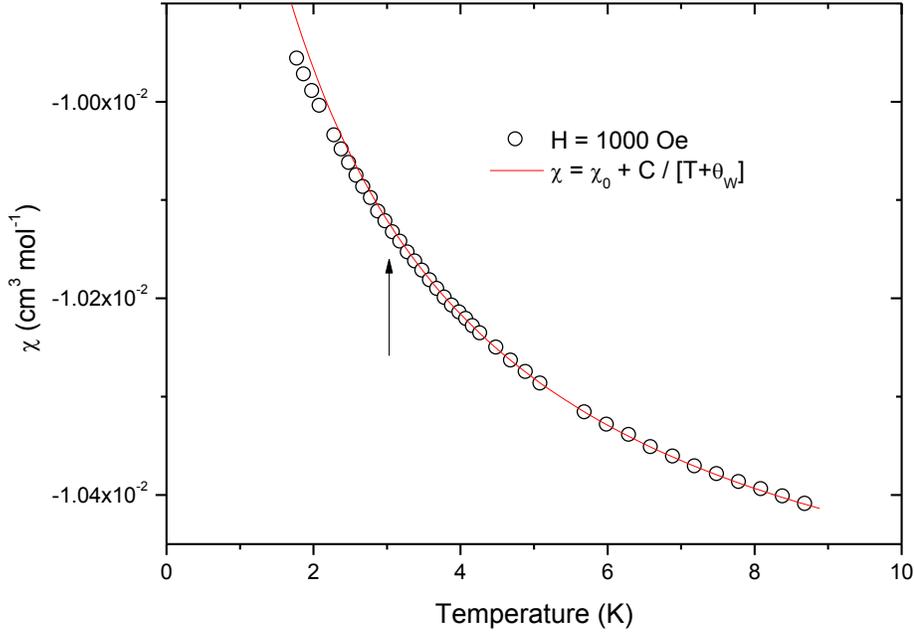

**Figure S3. Molar susceptibility of Li-FLG due to paramagnetic impurities**. The field-cooled molar susceptibility $\chi$ for H=1000 Oe is analyzed by using the Curie-Weiss law, $\chi = \chi_0 + C/(T+\theta_w)$, where $\chi_0$ is the temperature independent background that explains the large diamagnetic component. The best result is obtained from the least-squares-fit of the Curie-Weiss behavior when the Weiss temperature $\theta_w$ is 1.3 K and the Curie constant C is 0.0022 cm$^3$ K mol$^{-1}$. With an assumption that magnetic moment of each defect is $2\mu_B$, the concentration of defects is estimated to be less than 0.012 % per formula unit. Here, the Curie constant is $N(2\mu_B)^2\mu_0/k_B$, where N is the number of magnetic defects, $k_B$ is the Boltzman constant, $\mu_0$ is the magnetic permeability in free space, and $\mu_B$ is the Bohr magneton.

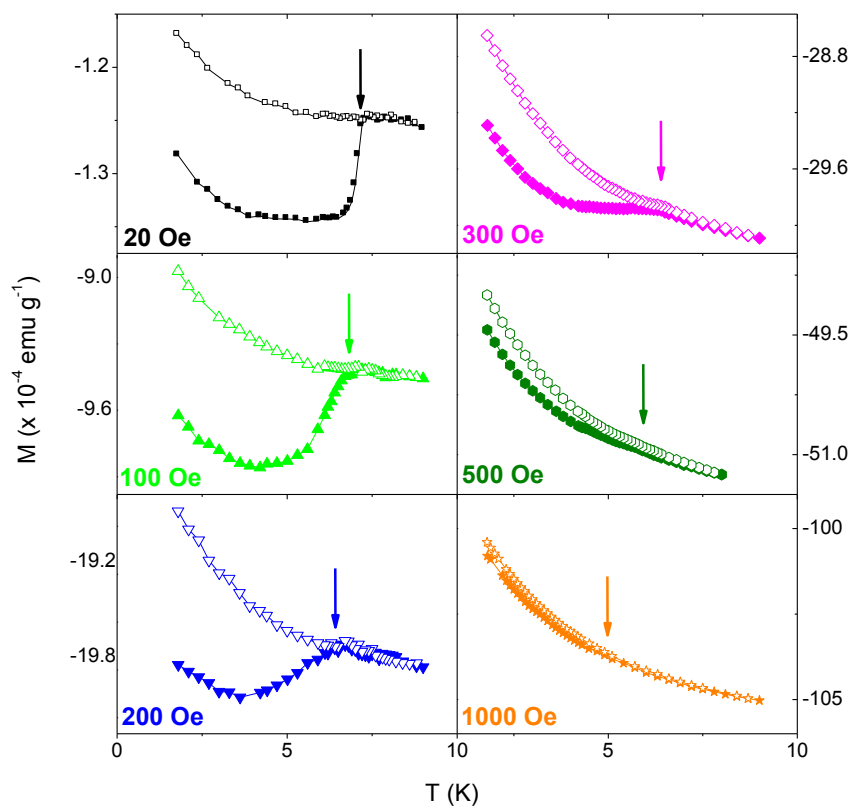

**Figure S4.** *M vs T* **of Li-intercalated FLG for several magnetic fields.** Zero-field-cooled (ZFC) and field-cooled (FC) magnetization data is plotted as a function of temperature and described by solid and open symbols, respectively. Arrows indicate onset of the superconducting phase transition temperature $T_c$. With increasing magnetic field, $T_c$ onset is gradually suppressed, while the diamagnetic signal becomes weaker. At 1000 Oe, the difference between ZFC and FC is almost negligible, which could be ascribed to the small grain size of the Li-FLG.

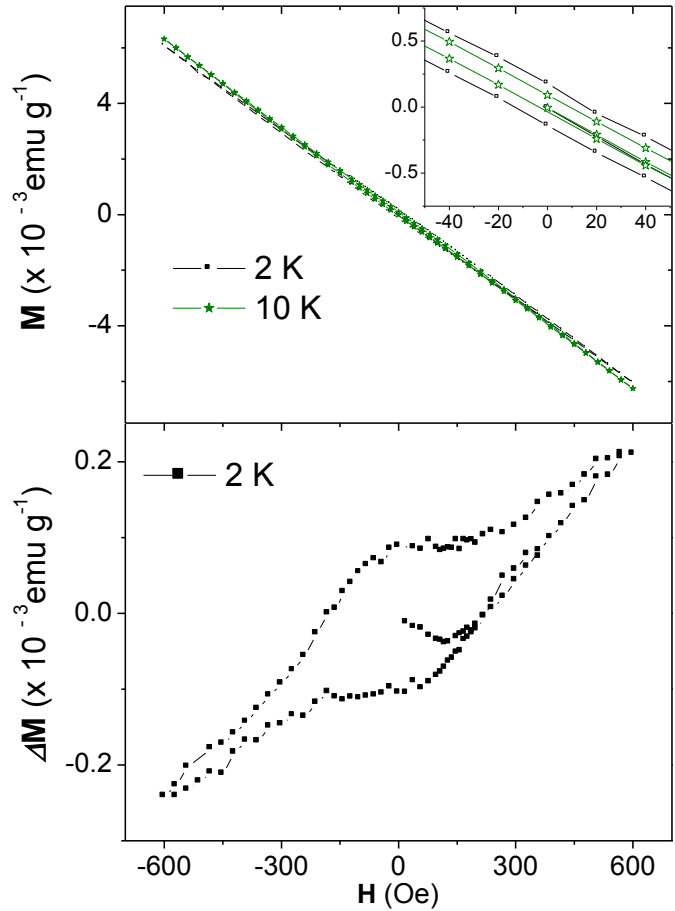

**Figure S5. Field dependence of the magnetization of the Li intercalated FLG. Top panel:** Magnetization of Li-FLG is shown as a function of magnetic field (*M-H* loop) for 2 (squares) and 10 K (stars). Inset is the blow-up of *M* near 0 Oe. **Bottom panel:** Δ*M* – *H* loop at 2 K is representatively plotted, where *M* at 2 K is subtracted by *M* at 10 K: Δ*M = M - M(10K)*. Here *M* at 10 K reflects the field dependence of *M* in the normal state of Li-FLG. We note that the magnetization crosses over zero into a positive value with increasing field, which could be ascribed to the presence of magnetic impurities that may arise from defects in the graphene. Field dependence of the magnetization is consistent with an increase in *M* with decreasing temperature.

**Supplementary References:**